\documentclass[aps,prb,twocolumn,superscriptaddress,showpacs]{revtex4}
\usepackage[T1]{fontenc}
\usepackage[latin9]{inputenc}
\usepackage{amsmath}
\usepackage{amssymb}
\usepackage{graphicx}
\usepackage{esint}
\usepackage{bm}

\renewcommand{\vec}[1]{\bm{#1}}

\DeclareMathAlphabet{\mathsfsl}{OT1}{cmss}{m}{sl}

\begin{document}

\title{Building topological device through emerging robust helical surface states}

\author{Zibo Wang}
\affiliation{International Center for Quantum Materials, Peking University, Beijing 100871, China}
\affiliation{Collaborative Innovation Center of Quantum Matter, Beijing 100871, China}

\author{Juntao Song}
\affiliation{Department of Physics, Hebei Normal University, Hebei 050024, China}

\author{Haiwen Liu}
\affiliation{International Center for Quantum Materials, Peking University, Beijing 100871, China}
\affiliation{Collaborative Innovation Center of Quantum Matter, Beijing 100871, China}

\author{Hua Jiang}
\thanks{\texttt{jianghuaphy@suda.edu.cn}}
\affiliation{Department of Physics, Soochow University, Suzhou 215006, China}

\author{X. C. Xie}
\affiliation{International Center for Quantum Materials, Peking University, Beijing 100871, China}
\affiliation{Collaborative Innovation Center of Quantum Matter, Beijing 100871, China}

\date{\today}

\begin{abstract}
We propose a nonlocal manipulation method to build topological devices
through emerging robust helical surface states in $Z_2=0$ topological systems.
Specifically, in a ribbon of $Z_2=0$ Bernevig-Hughes-Zhang (BHZ) model with finite-size effect,
if magnetic impurities are doped on the top (bottom) edge,
the edge states on the bottom (top) edge can be altered according to the strengths and directions of these magnetic impurities.
Consequently, the backscattering between the emerging robust helical edge states
and gapped normal edge states due to finite-size confinement is also changed,
which makes the system alternate between a perfect one-channel conductor and a perfect insulator.
This effect allows us to fabricate topological devices with high on-off ratio.
Moreover, it can also be generalized to 3D model
and more realistic Cd$_3$As$_2$ type Dirac semimetals.
\end{abstract}

\date{\today}

\maketitle

\section{Introduction}
The $Z_2=1$ quantum spin hall states and the consequent three-dimensional (3D) strong topological insulators (STI)
have attracted great interests since the success in their theoretical predictions and experimental observations.
\cite{M. Z. Hasan, C. L. Kane, M. Konig, H.-J. Zhang, D. Hsieh1, T. Hanaguri, Y. L. Chen}
One of the most important reasons is that there exist helical surface states characterized by a full insulating gap
and protected by the time-reversal symmetry.\cite{D. Hsieh2, Tong Zhang} Notably, those surface states are spin-momentum locked
and exhibit robustness against the nonmagnetic impurities, which makes $Z_2=1$ STI an ideal candidate
for energy-saving topological devices.

However, there still exist great challenges in the application of $Z_2=1$ STI due to their own intrinsic characteristics.
Specifically, as a topological device, its on and off status should be controlled,
which means the system can be alternated between a conductor and an insulator according to an external field.
Since the backscattering is suppressed with nonmagnetic impurities,
$Z_2=1$ STI is born a perfect conductor. Yet for this reason,
it is difficult to realize the off status in $Z_2=1$ STI.
By now, the most usual manipulation method is to dope magnetic impurities or apply an electromagnetic field,
\cite{Yoshinori Okada, L. Andrew Wray, Huichao Li, Xiaofeng Qian}
which all encounter the same problems.
First, both surfaces of the topological system should be doped in order to induce backscattering,
which also makes the system difficult to return to a conductor.
Second, since the backscattering only exists where the magnetic impurities or external fields are induced,
the surface states can only be controlled locally.
However, experimentally, topological insulators usually grow on a substrate,
which means any manipulation on a substrate surface becomes nearly impossible due to the difficulty of doping there.
Consequently, a nonlocal method to control the surface states close to the substrate
in order to build topological devices becomes rather an important topic.

Not long ago, theorists proposed a new kind of $Z_2=0$ topological system.\cite{Hua Jiang, Huaiming Guo, Bart de Leeuw, T. Fukui}
Considering finite-size effect, the $Z_2=0$ topological system with proper size inherits properties of $Z_2=1$ STI,
besides unique emerging robust helical surface states due to the finite-size confinement.
Here, the words ``robust'' and ``emerging'' mean those helical surface states are robust against nonmagnetic disorder
and only exist in a finite-size system compared to the traditional one.
Recently, Zhu et al. found the sign of these $Z_2=0$ emerging robust helical surface states
in epitaxial Bi(111) thin films.\cite{Kai Zhu} Moreover, in contrast to $Z_2=1$ STI, there also exists an unexpected phenomenon
that the weak anti-localization peak suddenly disappears after doping magnetic Co on the top surface of thin Bi film.
This experiment implies that the surface states on the bottom of thin $Z_2=0$ topological system
can be manipulated through a nonlocal method on the top surface, which may
have great significance in building topological devices.

In this work, we propose a new nonlocal method to build topological devices through
the emerging robust helical edge or surface states in $Z_2=0$ topological systems.
First, we demonstrate the method in a narrow
ribbon described by 2D $Z_2=0$ anisotropic BHZ model.
Without magnetic doping,
we find that there exist one pair of emerging robust helical edge states
and one pair of gapped normal edge states due to the finite-size confinement.
Therefore, the system exhibits conducting characteristics. However, if
the top (bottom) of the 2D BHZ model is doped with magnetic impurities,
the gapped normal edge states on the bottom (top)
gradually become gapless with increasing magnetization.
Thus, for moderate magnetization on the top (bottom) edge,
the system remains good conductor since there is no backscattering on the bottom (top) edge.
When magnetization becomes strong sufficiently, the Anderson disorder can cause strong backscattering on both edges,
resulting insulating characteristics for the system.
Using the Landauer-B\"{u}ttiker formula, we verify the above physical pictures by the two-terminal conductance $G$,
which starts from $2e^2/h$, then maintains at the value of $e^2/h$, and finally falls to zero.
This novel phenomenon enables us to nonlocally manipulate the on and off status in a $Z_2=0$ topological system,
thus it can also be applied to build new topological devices.
Our proposal is also extended to 3D $Z_2=0$ topological systems, such as Wilson-Dirac model.
And finally, we further discuss the realization of this proposal in Cd$_3$As$_2$ type Dirac semimetals.

The rest of this paper is organized as follows. In Sec.II, we first numerically demonstrate this
nonlocal engineering method in a 2D anisotropic BHZ
model and 3D anisotropic Wilson-Dirac model.
Then, we exhibit two detailed plans to build topological devices with $Z_2=0$ topological systems.
Next, in Sec.III, we generalize this
method to more realistic Cd$_3$As$_2$ Dirac semimetal materials.
Finally, a conclusion is presented in Sec.IV.

\section{2D BHZ model and 3D Wilson-Dirac model}

We first take the anisotropic BHZ model in square lattice as an example.\cite{B. A. Bernevig, X.-L. Qi}
The four-band tight-binding Hamiltonian in the momentum representation reads:
\begin{eqnarray}
    \mathcal{H}(\vec{k}) & = & \left(
                             \begin{array}{cc}
                             H(\vec{k}) & 0\\
                             0 & H^*(-\vec{k})
                             \end{array}
                         \right) \nonumber\\
              H(\vec{k}) & = & \tau_z(m-m_x-m_y+m_x\cos k_x+m_y\cos k_y) \nonumber\\
                         &   & +\tau_xv_x\sin k_x+\tau_yv_y\sin k_y
\end{eqnarray}
where $m$ determines the band gap, $v_{x,y}$ reflects the Fermi velocity, and $m_{x,y}$ represents the
hopping amplitude between nearest-neighbor sites along the x, y directions, respectively.
$\tau_{x,y,z}$ are Pauli matrices representing different orbits.
During our calculation, the values of these parameters are chosen as below:
$m=1.64, m_x=0.8, m_y=1.2, v_{x,y}=3$.
We consider the ribbon geometry, and the width along y direction is chosen as $L_y=30a$ ($a$ is the lattice constant)
in order to induce finite-size effect.

In this work, our plan of magnetic doping is to dope a soft-magnetic stripe on one edge
or surface of the sample. Therefore, the strengths and directions of the magnetic impurities
are entirely determined by external magnetic field,
and we consider the effect of magnetic impurities in x direction by an adding term
$M_x \sigma_x \otimes \tau_0$ on $y=0$ edge of the ribbon.
Actually, even though it will be difficult to accurately control the magnetization
strength $M_x$ of the soft-magnetic stripe in experiment,
an external magnetic field rotating in x-z direction will also work.
Because $M_x$ is determined by $M\cos \theta$,
and the existence of $M_z$ has little influence on the final result.

\begin{figure}[h]
\includegraphics [width=\columnwidth, viewport=0 0 975 508, clip]{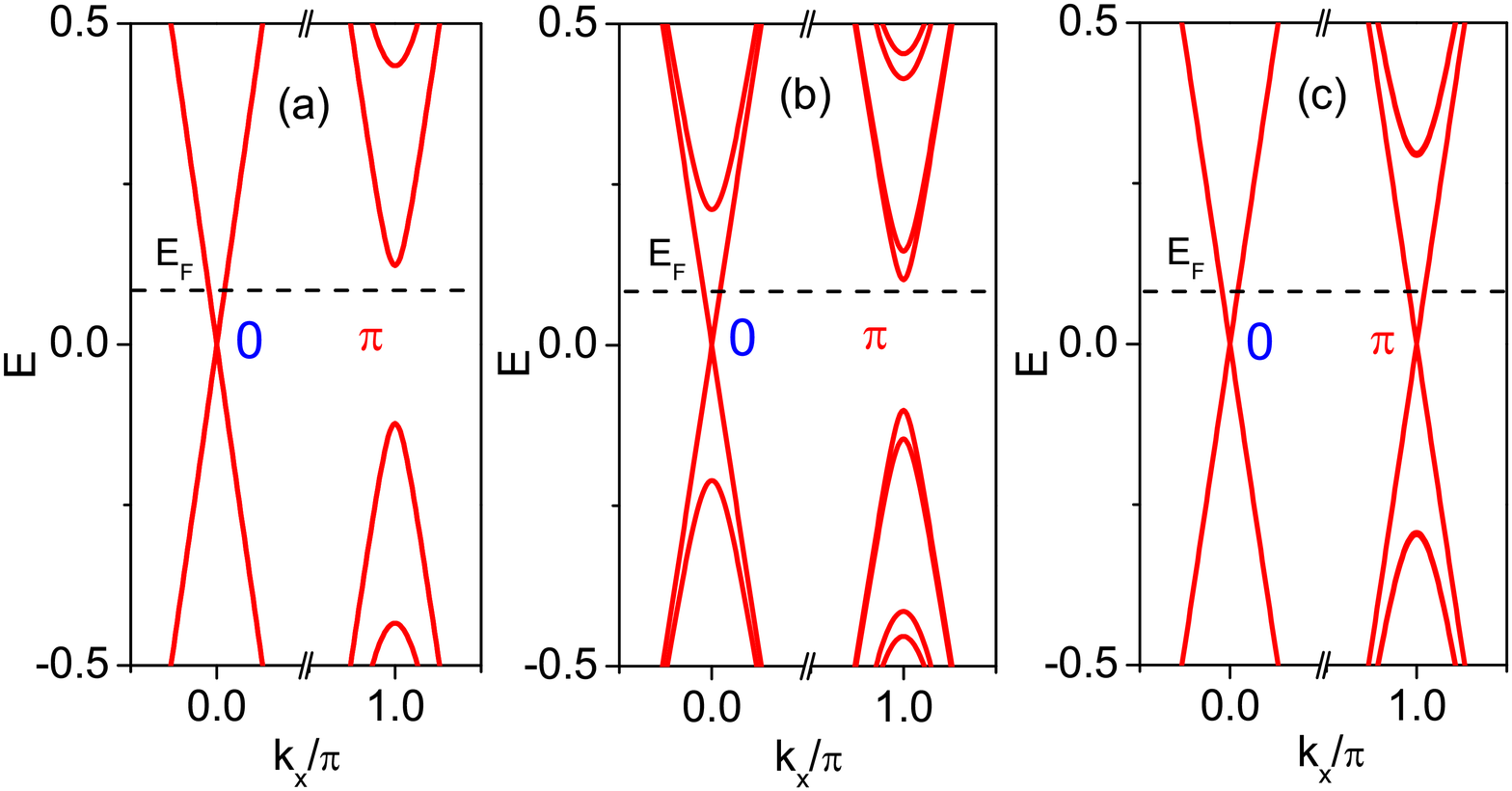}
\includegraphics [width=\columnwidth, viewport=0 0 1013 377, clip]{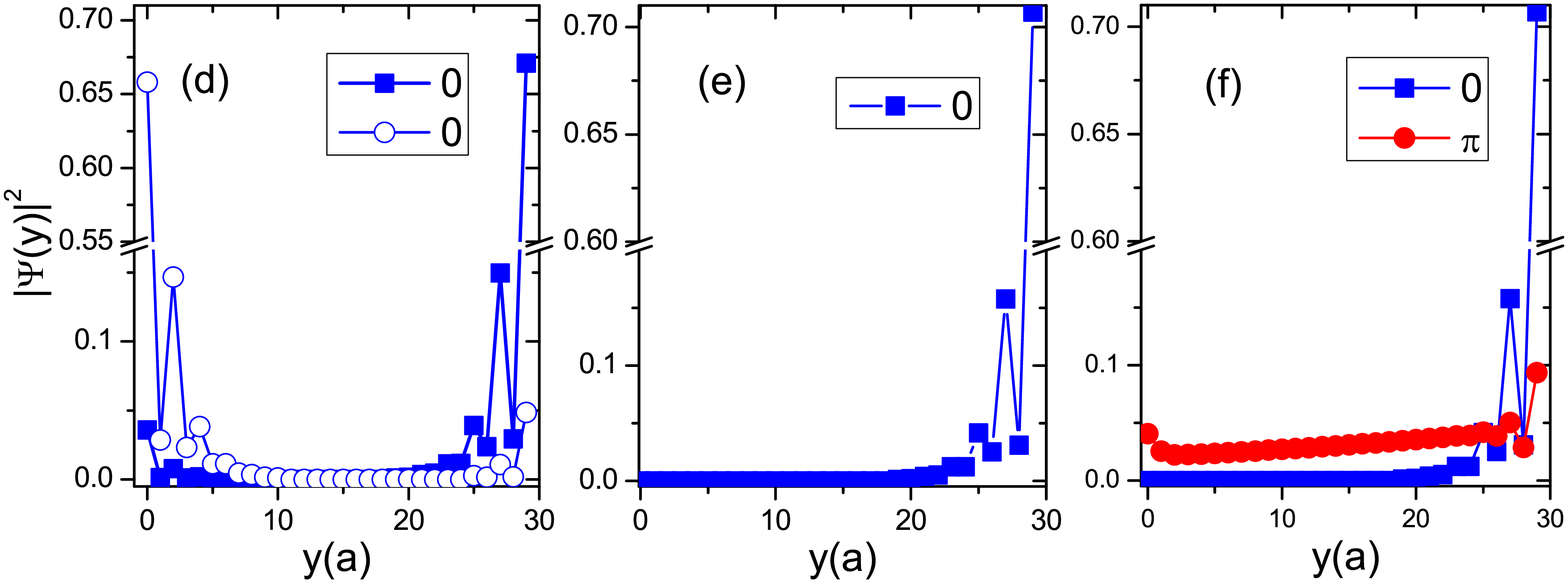}
\caption{(Color online)
(a)-(c) stand for calculated energy bands in 2D BHZ model with three different strengths of magnetic impurities doped in x direction:
$M_x=0, 1$ and 2.1, respectively. (d)-(f) represent corresponding probability density
distributions of the edge states for (a)-(c) at $E_F=0.08$,
respectively. Only the left propagating states are plotted here.}\label{bulkbandsevolution}
\end{figure}

Without magnetic impurities, the energy band of the system is drawn in Fig.1(a),
which is double degenerated due to the spin degeneration.
There exist two pairs of gapless edge states and two pairs of gapped edge states
locating at $k_x=0$ and $k_x=\pi$, respectively.
When the Fermi energy is set $E_F=0.08$ as the black dashed line shows in Fig.1(a),
the corresponding probability density $|\Psi(y)|^2$ is drawn in Fig.1(d).
It is obvious that the gapless edge states around $k_x=0$ mostly locate at two edges, representing the emerging robust helical edge states.
Consequently, the edge states around $k_x=\pi$ represent the hybridized gapped normal edge states due to the finite-size confinement.

The energy band with magnetic impurity strength $M_x=1$ is shown in Fig.1(b).
We find that one pair of the previously double degenerated edge states around $k_x=0$ opens a small gap,
while the other one remains close.
Moreover, one of the originally double degenerated energy gaps around $k_x=\pi$ becomes smaller.
In Fig.1(e), we find that the only existing edge states are the
emerging robust helical ones locating at $y=30a$.

Continuing increasing the impurity strength, the above tendency exhibits much more explicitly.
If the impurity strength reaches as strong as $M_x=2.1$,
the smaller gap in Fig.1(b) around $k_x=\pi$ becomes nearly gapless in Fig.1(c).
In Fig.1(f), the corresponding probability density $|\Psi(y)|^2$ tells us that the only existing edge states are those locating around $y=30a$.
Specifically, the emerging robust helical edge state colored by blue mostly locates around $y=30a$,
while the normal edge state colored by red extends to inner space,
which means it still exhibits properties of bulk states.

\begin{figure}[h]
\includegraphics [width=\columnwidth, viewport=0 0 594 364, clip]{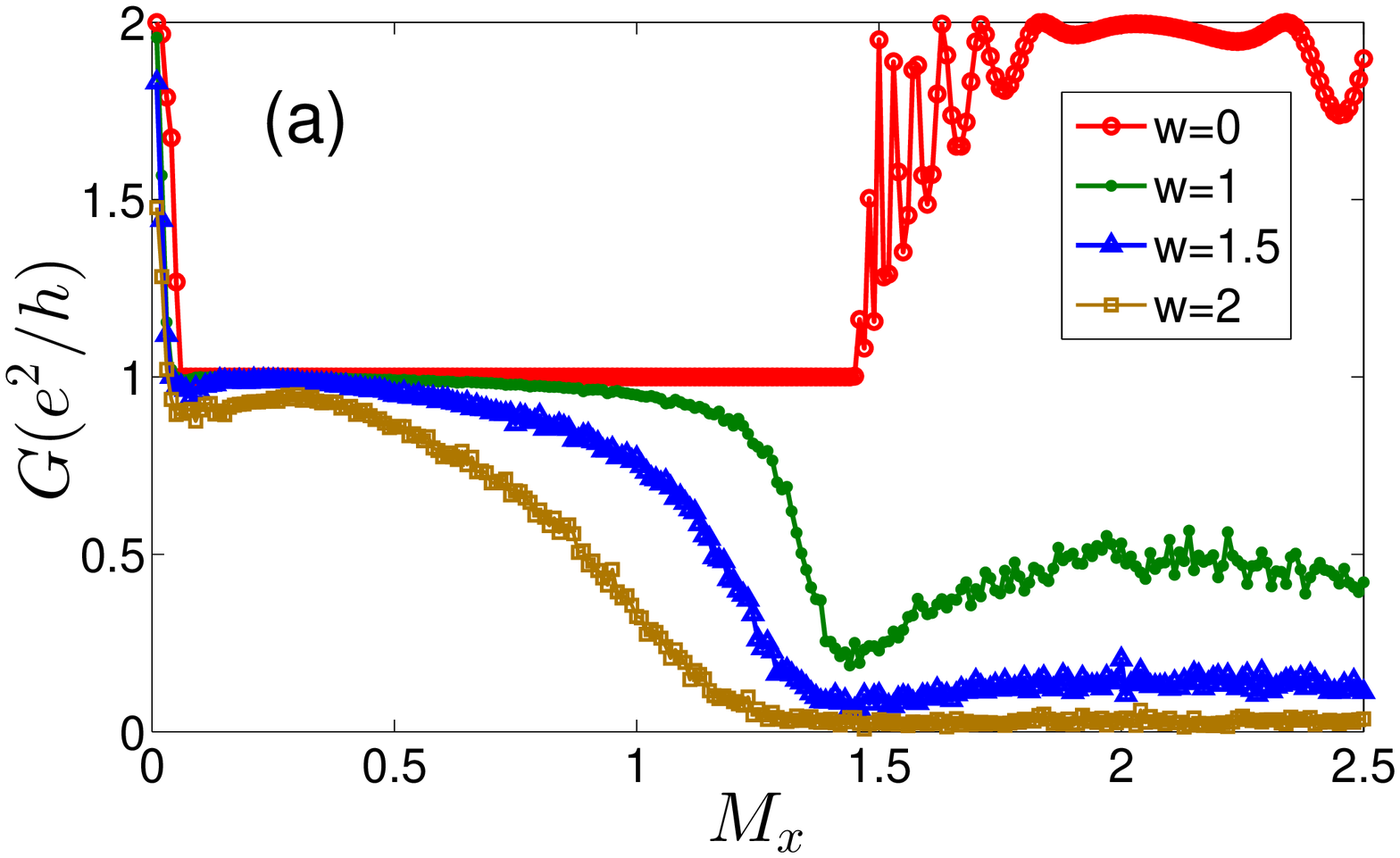}
\includegraphics [width=\columnwidth, viewport=0 0 1785 741, clip]{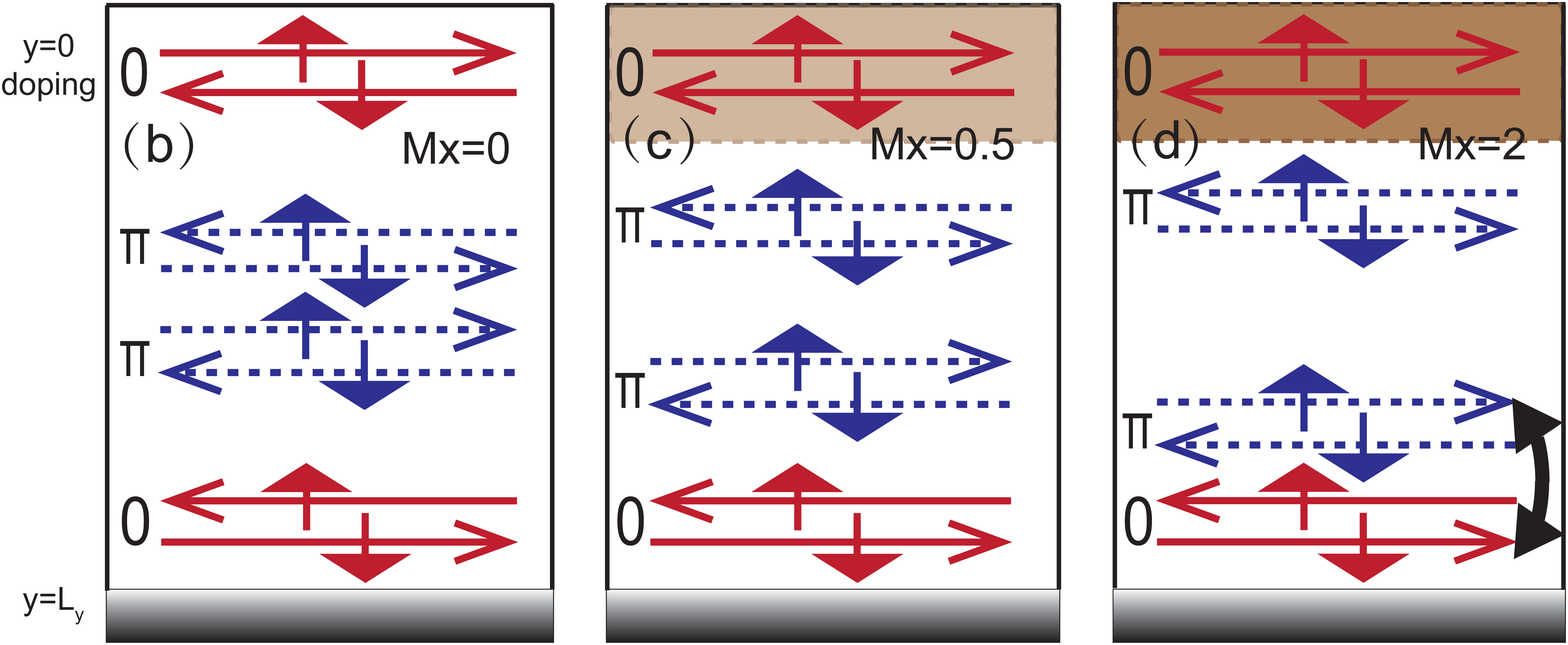}
\includegraphics [width=\columnwidth, viewport=0 0 783 320, clip]{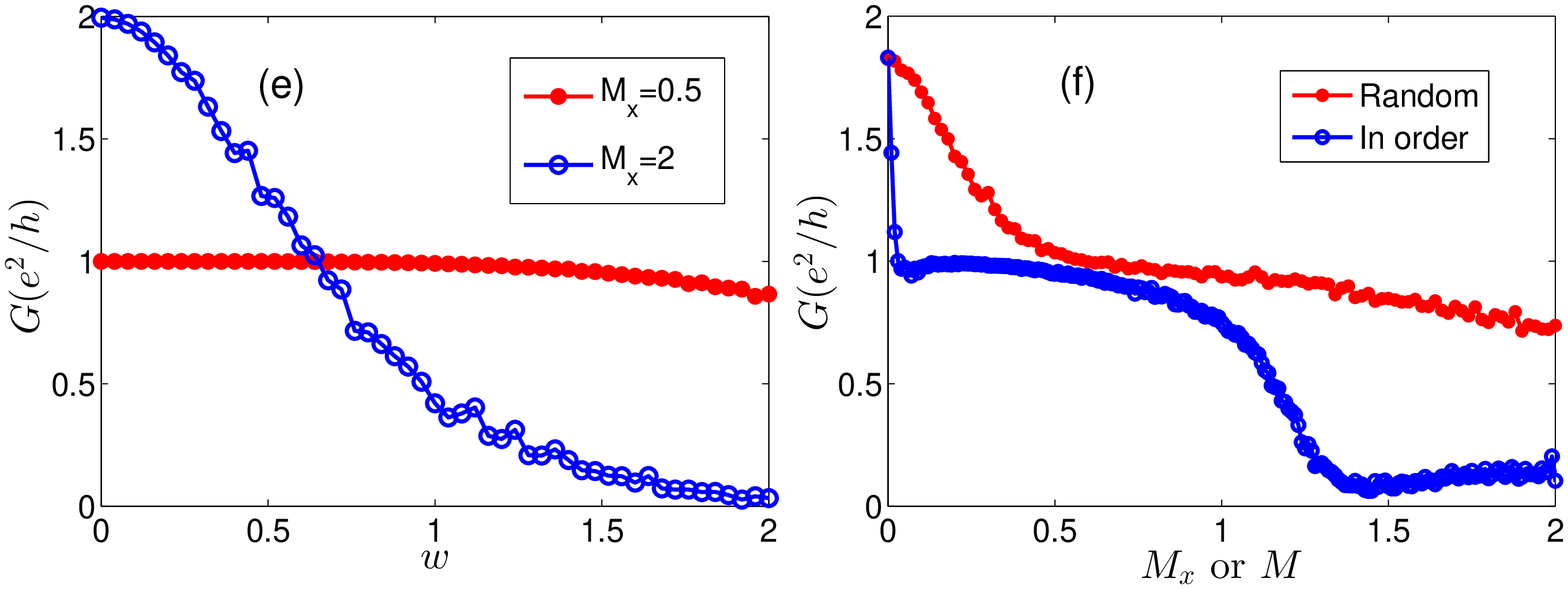}
\caption{(Color online)
(a) is the conductance $G$ vs $M_x$ for different Anderson disorder strength $w$ in a two-terminal system.
The parameters $E_L=E_R=2.3$ for left and right leads, $E_C=0.03$ for central part.
(b)-(d) are the schematic diagrams corresponding to the three characteristic stages of the conductance
in (a) during the process of doping magnetic impurities.
(e) is the conductance $G$ vs $w$ with fixed $M_x$.
(f) is the conductance $G$ vs $M_x$ or $M$ for magnetic impurities doped in order or randomly.}\label{bulkbandsevolution}
\end{figure}

In order to quantitatively analysis the physical pictures of energy bands,
we further study the transport properties of this system,
and calculate the conductance $G$ of this model with nonmagnetic disorder in a two-terminal system.
The nonmagnetic disorder exists inevitably in the realistic system
and is modeled by Anderson disorder with random potential uniformly distributed in $[-w/2,w/2]$, where $w$ is the disorder strength.
The distance between the two leads is $3000a$ and the Fermi energy of the central region is $E_C=0.03$.
In order to simulate metallic leads at the two terminals, their widths along y direction and Fermi energies
are chosen as $N_L=N_R=200a$ and $E_L=E_R=2.3$, respectively.

We first show how the conductance varies with the strength of magnetic impurities $M_x$ in Fig.2(a).
For the red line without Anderson disorder,
the conductance undergoes a transition from $G=2e^2/h$ to $e^2/h$, then back to $2e^2/h$.
The second plateau of $G=2e^2/h$ shakes vigorously, because the normal edge states on $y=30a$
still have properties of bulk states as shown by the red line in Fig.1(f).
The whole transition processes can be understood with the help of the schematic diagrams from Fig.2(b) to 2(d).
First, as shown in Fig.2(b), due to the finite-size effect,
there exist one pair of emerging robust helical edge states and one pair of gapped normal edge states,\cite{Hua Jiang}
while the gapped normal ones have no contribution to the conductance.
Therefore, the conductance exhibits $G=2e^2/h$ at the beginning.
Then, in Fig.2(c), if we slightly dope a little amount of magnetic impurities in x direction on $y=0$ edge,
which will induce the backscattering between the
counter propagating emerging robust helical edge states on $y=0$,
only the emerging robust helical edge states on $y=L_y$ have contribution
to the conductance. Thus, the conductance falls to $G=e^2/h$ fast and then remains unchanged.
At last, in Fig.2(d), when $M_x$ is strong sufficiently,
the gapped normal edge states on $y=0$ owns a relatively large gap,
and the coupling between the gapped normal edge states on $y=0$ and $y=L_y$ is really weak.
Therefore, the originally gapped normal edge states on $y=L_y$ become gapless
and could make contribution to the conductance.
As a result, the conductance gets back to $G=2e^2/h$.

However, considering the inevitable impurities in reality,
the system shows exotic transport properties, which is the central result
of this paper. If the strength of Anderson disorder is not zero,
as shown by the green, blue and brown lines in Fig.2(a), the conductance never returns to $G=2e^2/h$,
but continues falling to nearly zero with strong $M_x$.
This difference can be explained by the backscattering between the emerging robust
helical edge states and the normal ones on $y=30a$ caused by Anderson disorder.
The existence of this backscattering is guaranteed by the physical
meaning of $Z_2$ topological invariant: gapless helical edge states with even number are not
robust against disorder.
Besides, for the green line, when the impurity strength approaches $M_x=1.5$,
we find the conductance exhibits a minimum value during the process of decreasing.
This critical point is actually also the transition point from $G=e^2/h$ to $2e^2/h$ without disorder,
which means the Fermi energy just touches the bottom of the closing energy band on $k_x=\pi$.
In this case, the backscattering in gapped $k_x=\pi$ channel is permitted.
Therefore, there exist more backscattering channels shown in Fig.2(d), and the
backscattering between the emerging robust helical edge states and the gapped normal ones on $y=30a$
is somehow enhanced. As a result, the conductance at this point exhibits smaller value than
the surrounding points.

Next, let's fix the magnetic impurity strength $M_x$,
and study how the the conductance $G$ varies with Anderson disorder strength $w$.
In Fig.2(e), it is obvious that the conductance with $M_x=0.5$
exhibits stronger robustness than that with $M_x=2$, which is consistent with Fig.2(a).
Moreover, for instance, if the disorder strength is chosen as $w=2$,
the ``on'' status of the red line with $M_x=0.5$ exhibits $G\approx 0.87e^2/h$,
while the ``off'' status of the blue line with $M_x=2$ exhibits $G\approx 0.03e^2/h$.
Therefore, the on-off ratio can be achieved about $30:1$ in this case.
Till now, we find that by altering the strength of magnetic impurities on one edge, a $Z_2=0$ BHZ model
with emerging robust helical edge states can be tuned from a perfect one-channel conductor
to a perfect insulator, which can be applied to build topological devices.

However, in some cases, it may not be convenient to alter the external field,
and the strength of magnetic impurities $M_x$ is fixed.
Therefore, let's consider a situation that the direction of doped magnetic impurities is not ordered but in a random arrangement.
For simplicity, we just assume that $\vec{M}$ is either in x or z direction.
During our calculation, $M_x^2+M_z^2=M_{total}^2$, $M_x\in[-M_{total},M_{total}]$, and $M_{total}\in[0,M]$.
The red line of Fig.2(f) shows how the conductance varies with magnetic impurity strength $M$ when $w=1.5$.
To have a better comparison, a similar $w=1.5$ situation with magnetic impurities ordered in x direction is also plotted by a blue line.
As we can see, when $M=2$,
the conductance of the red line still maintains at $G\approx 0.8e^2/h$ compared with $G\approx 0.1e^2/h$ of the blue line when $M_x=2$,
which can be regarded as the on and off status of the current.
In experiment, this kind of random can be realized in many ways, such as by heating and so on.
Therefore, topological devices built by $Z_2=0$ 2D BHZ model with emerging robust helical edge states
can also be tuned from a perfect one-channel conductor to a perfect insulator
by altering the directions of magnetic impurities on one edge.

\begin{figure}[h]
\includegraphics [width=\columnwidth, viewport=0 0 782 555, clip]{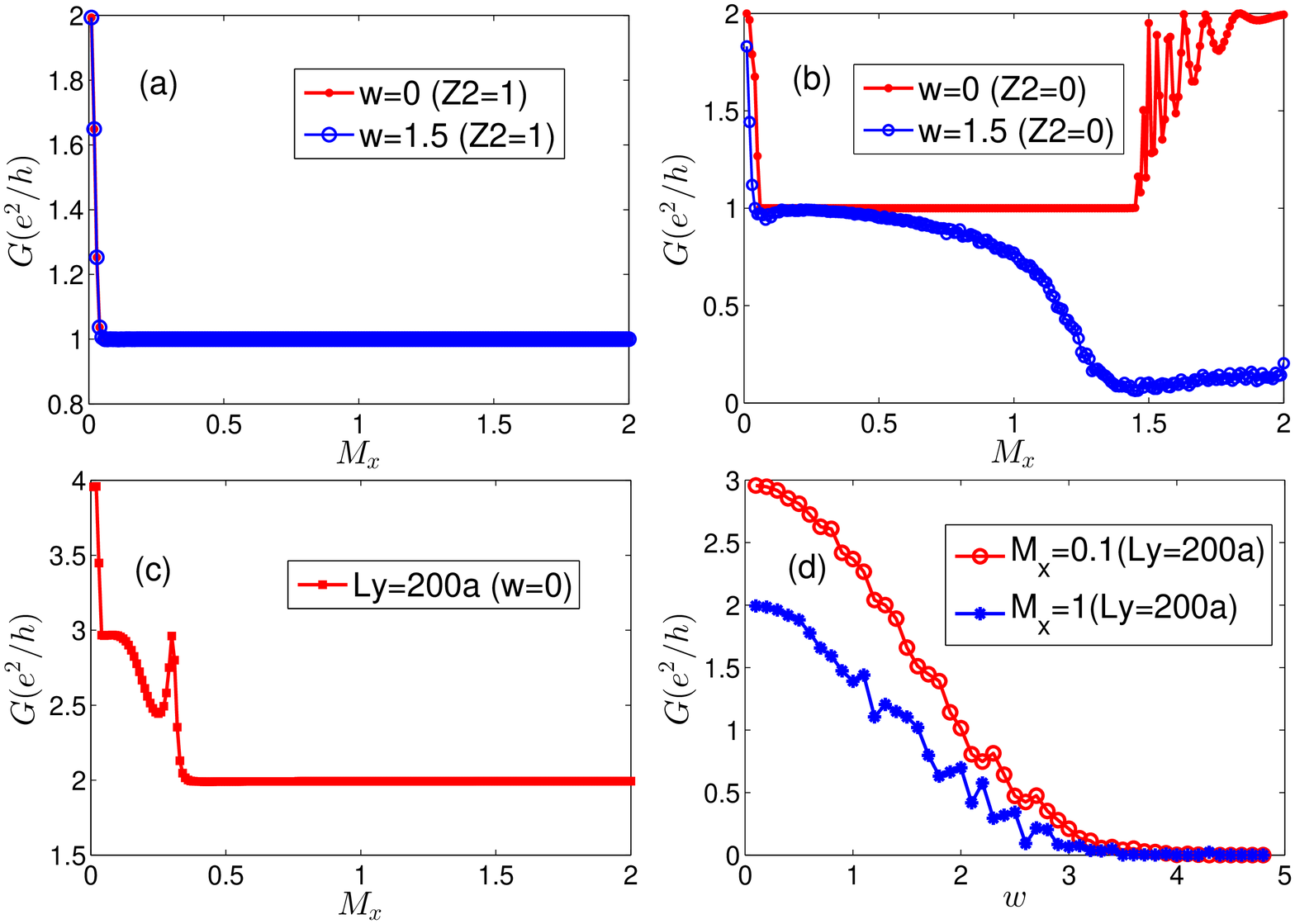}
\caption{(Color online)
(a): The conductance $G$ vs $M_x$ for $Z_2=1$ STI with $m=1$ in BHZ model.
(b): The conductance $G$ vs $M_x$ for $Z_2=0$ topological system with $m=1.64$ in BHZ model.
(c): The conductance $G$ vs $M_x$ for $L_y=200a$ without finite-size effect in BHZ model.
(d): The conductance $G$ vs disorder strength $w$ in BHZ model for $L_y=200a$ without finite-size effect.
In (c) and (d), all parameters of the Hamiltonian are chosen the same as Fig.2(a) except $L_y$.}\label{bulkbandsevolution}
\end{figure}

For further showing the advantages of the above proposals,
we also consider a 2D $Z_2=1$ BHZ model by changing the parameter $m$ in Eq(1) from 1.64 to 1
and $L_y$ still maintains $30a$.
As shown in Fig.3(a), no matter whether there exists Anderson disorder or not, the conductance
never falls to zero, but keeps $G=e^2/h$ all the time.
Compared with the $Z_2=0$ topological system in Fig.3(b), the $Z_2=1$ STI can not turn off the current,
which means it is not suitable to build topological devices.
Moreover, we also calculate a $Z_2=0$ 2D BHZ model without finite-size effect by expanding $L_y$ from $30a$ to $200a$.
The red line in Fig.3(c) represents how the conductance $G$ varies with magnetic impurity strength $M_x$ without Anderson disorder,
and there exists two characteristic quantum plateaus $G=3e^2/h$ and $G=2e^2/h$.
In Fig.3(d), we show how the conductance on these two plateaus varies with Anderson disorder strength $w$.
The red line and the blue line represent $M_x=0.1$ and $M_x=1$, respectively.
Unlike Fig.2(e),
the main feature of Fig.3(d) is that the conductances of these two lines start decreasing at the beginning, and
fall to zero nearly with the same $M_x$. This feature tells us that
one cannot get high on-off ratio for any disorder strength $w$ or magnetic impurity strength $M_x$.
The reason is that there are always two pairs of gapless helical edge states around $y=L_y$,
and the Anderson disorder can cause the backscattering between them
(the helical edge states around $k_x=0$ are no longer emerging robust ones now).
Therefore, if we want to switch on or off the current in order to build topological devices
just by tuning the strength of doped magnetic impurities on one edge,
the existence of $Z_2=0$ emerging robust helical edge states due to finite-size confinement is the necessary condition.

\begin{figure*}[htbp!]
\scalebox{2}{\includegraphics [width=\columnwidth, viewport=0 0 1756 519, clip]{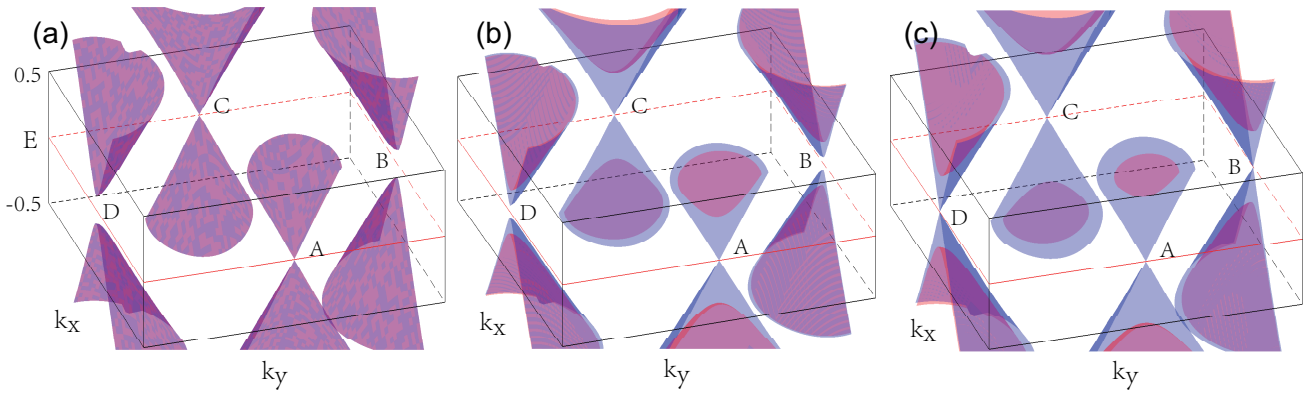}}
\caption{(Color online)
(a)-(c) correspond to the surface energy bands of Wilson-Dirac model with different strengths of magnetic
impurities: $M_z=0,1$ and $2$, respectively.
As $M_z$ increases, one of the two double degenerated gapless surface states on $A(C)$ opens a small gap,
while one of the two double degenerated gapped surface states on $B(D)$ gradually becomes gapless.
}\label{bulkbandsevolution}
\end{figure*}

Similar phenomenon also exists in 3D $Z_2=0$ topological systems,
and we take the anisotropic Wilson-Dirac type model as an example.
\cite{K.-I. Imura, R. S. K. Mong, Z. Ringel, K. Kobayashi, Y. Yoshimura, C.-X. Liu}
The Hamiltonian in a cubic lattice reads:
\begin{eqnarray}
    \mathcal{H}(\vec{k}) & = & m(\vec{k})\sigma_0\otimes\tau_z+\sum_\alpha v_\alpha \sin k_\alpha\sigma_\alpha\otimes\tau_x \nonumber\\
              m(\vec{k}) & = & m + \sum_\alpha m_\alpha(\cos k_\alpha-1)
\end{eqnarray}
where $\sigma$ are Pauli matrices in spin space, and $\alpha=x,y,z$.
Parameters $\tau, m, v_\alpha, m_\alpha$ have the same meanings as the
2D model of Eq(1). The width along z direction $L_z$ is chosen as $L_z=15a$,
and the magnetic impurities are doped in z direction on $z=0$ side.
During the calculation, related parameters are chosen as below: $m=2.26, m_x=0.9, m_y=1.1, m_z=0.8, v_{x,y,z}=1.5$.
In Fig.4(a)-(c), we show how the 3D energy band varies with the strength
of doped magnetic impurities $M_z=0, 1$ and $2$, respectively.
There exist two nonequivalent Dirac points $A(C)$ and $B(D)$,
and the blue and pink curves represent the two energy bands nearest to the Dirac points.
In Fig.4(a), the energy bands locating around $A(C)$ are gapless, because they are the emerging robust helical surface states.
However, the appearance of the gaps on $B(D)$ tells us these surface states are the gapped normal ones due to finite-size confinement.
In Fig.4(b), $M_z=1$, we find that the pink curves on $A(C)$, which were gapless at first,
now open a small gap, because there exists the backscattering
between the counter propagating emerging robust helical surface states
on $z=0$ induced by the magnetic impurities. At last, in Fig.4(c), when the
strength of magnetic impurities is strong sufficiently,
all blue curves at $A(C)$ and $B(D)$ become gapless.
Briefly, Dirac points $A(C)$ and $B(D)$ correspond to $k_x=0$ and $k_x=\pi$ in 2D BHZ model, respectively.
And all the physical pictures of Fig.4(a)-(c) are exactly the same as those shown by Fig.1(a)-(c), respectively.

By now, we have afforded two kinds of nonlocal methods
to build topological devices in $Z_2=0$ topological systems.
The first one is to alter the strengths of the doped magnetic impurities.
Specifically, we first dope a soft-magnetic film at the top
surface of a thin $Z_2=0$ topological system grown on a substrate.
Then, if we only apply a weak magnetic field on the soft-magnetic film,
the whole device is a good conductor.
However, if the soft-magnetic film is magnetized strong sufficiently,
the whole device turns into an insulator.
The second method is to alter the directions of doped magnetic impurities.
If the doped magnetic film is magnetized in order with moderate strength, the whole device is an insulator.
However, if the magnetic direction of doped magnetic film is in a random arrangement,
such as by heating and so on, the whole system gets back to a conductor.
Since the above manipulation methods are all nonlocal ones on one edge or surface of the sample,
it exhibits great advantages to build topological devices through emerging robust helical surface states in $Z_2=0$ topological systems
compared with traditional $Z_2=1$ STI.

\section{Extension to real materials}

Based on above two simple models,
we have in principle shown the nonlocal manipulation methods to build topological devices
with emerging robust helical surface states in $Z_2=0$ topological systems.
In this section, let's further extend our proposal to more realistic materials.
Though there are several 3D realistic material candidates,\cite{Kai Zhu, Xiao Li}
we restrict our discussion on 2D cases to avoid the huge computational requirements.

Motivated by the recent theoretical prediction and experimental realization of Dirac semimetal Cd$_3$As$_2$,
\cite{Zhijun Wang, Madhab Neupane, Sergey Borisenko}
we consider a Cd$_3$As$_2$ stripe whose effective low energy Hamiltonian $H(\vec{k})$ reads as:\cite{Zhijun Wang}
\begin{eqnarray}
    H(\vec{k})  = \epsilon_0(\vec{k})+ \left(
                             \begin{array}{cccc}
                             M(\vec{k}) & Ak_+ & Dk_- & B^*(\vec{k})\\
                             Ak_- & -M(\vec{k}) & B^*(\vec{k}) & 0\\
                             Dk_+ & B(\vec{k}) & M(\vec{k}) & -Ak_-\\
                             B(\vec{k}) & 0 & -Ak_+ & -M(\vec{k})
                             \end{array}
                         \right)
\end{eqnarray}
where $\epsilon_0(\vec{k})=C_0+C_1k^2_z+C_2(k^2_x+k^2_y)$, $k_\pm=k_x\pm ik_y$,
and $M(\vec{k})=M_0-M_1k^2_z-M_2(k^2_x+k^2_y)$ with parameters $M_0,M_1,M_2<0$.
Since $B(\vec{k})=(\alpha k_z+\beta D)k_z^2$, $B(\vec{k})$ can be neglected compared
with $Dk_\pm$ term if we only consider the expansion up to $O(k^2)$.\cite{Zhijun Wang, Sangjun Jeon}

\begin{figure}[h]
\includegraphics [width=\columnwidth, viewport=0 0 1007 420, clip]{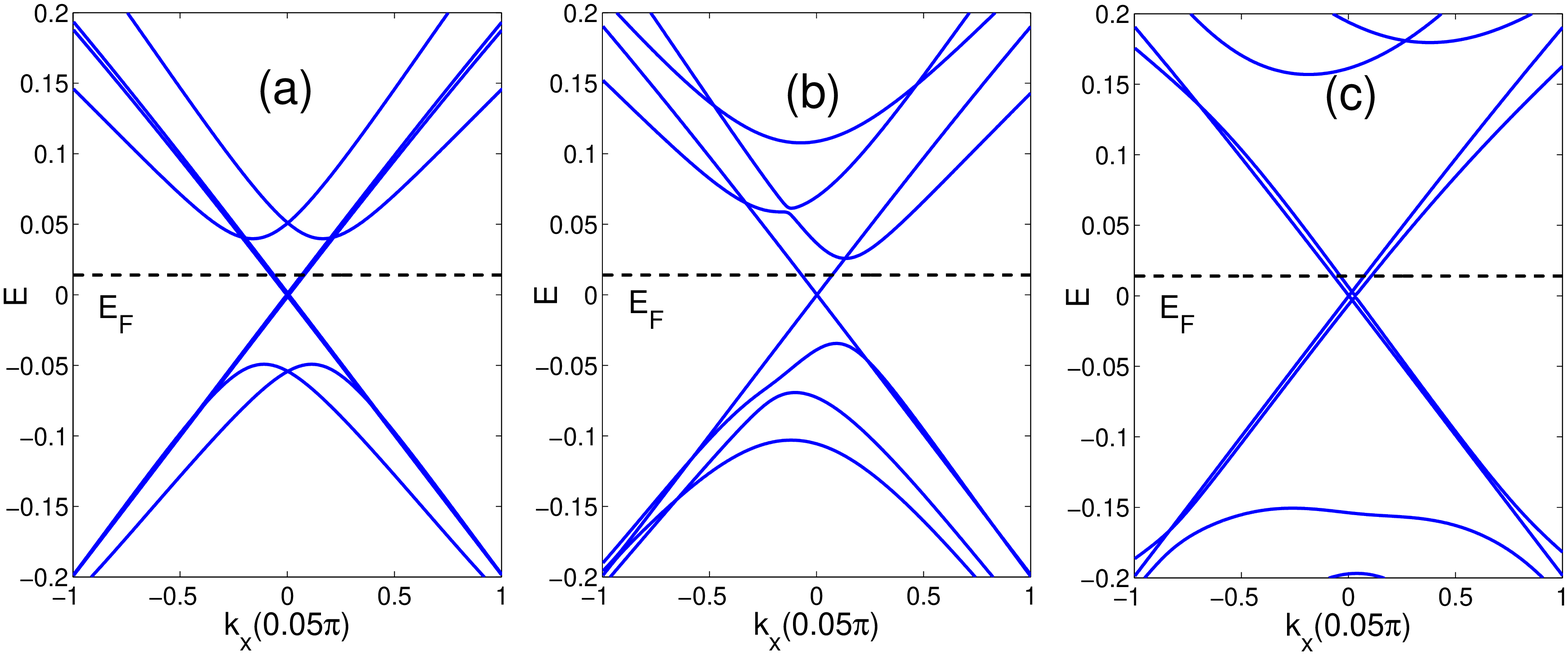}
\includegraphics [width=\columnwidth, viewport=0 0 686 238, clip]{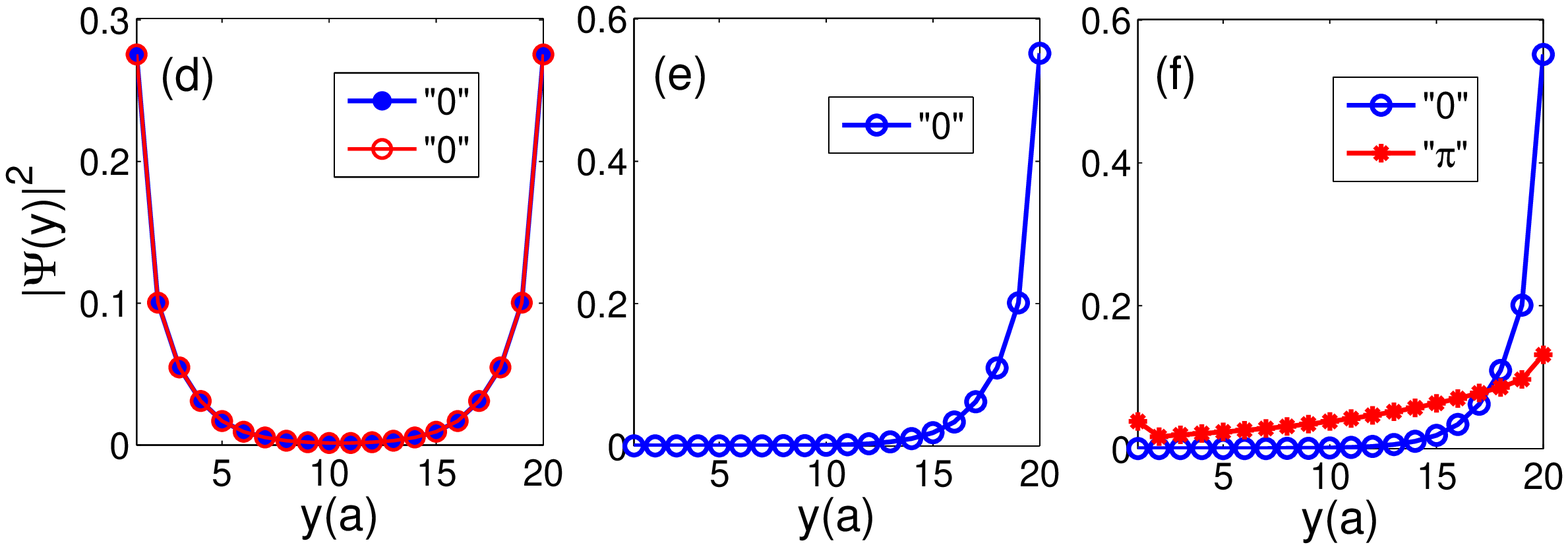}
\caption{(Color online)
(a)-(c) stand for calculated energy bands in Cd$_3$As$_2$ for three different strengths of
magnetic impurities doped in x direction: $M_x= 0,0.2$ and $0.9$, respectively.
(d)-(f) are the probability density distributions which has been integrated along $z$ axis
for (a)-(c) at $E_F$ = 0.014, respectively. In (d)-(f), only the left propagating states are plotted.}\label{bulkbandsevolution}
\end{figure}

In this paper, $C_0,C_1$ and $C_2$ are chosen zero, which can simplify our numerical calculation
but has no influence on the physical properties of this model.
In fact, main results got from this approximation satisfy realistic Cd$_3$As$_2$ materials,
though they are not quite true in experiment.
Other related parameters are chosen as below: $M_0=-0.44, M_{1,2}=-0.5, A=1.2, D=0.55$.
Considering 3D stripe geometry in which $L_z=6a$ and $L_y=20a$,
the above $\vec{k}\cdot \vec{p}$ Hamiltonian can be rewritten in tight-bind form, and
the energy bands of Cd$_3$As$_2$ are drawn in Fig.5(a)-(c).
Compared with Fig.1(a)-(c), the two Dirac points $k_x=0$ and $k_x=\pi$ in BHZ model
both locate at $k_x=0$ with two inverted bands.\cite{Zhijun Wang}
In Fig.5(a), there is no doped magnetic impurity. The two pairs of gapless surface states
represent the emerging robust helical ones, and the other two with a visible gap represent
the gapped normal ones.
Taking the Fermi energy as $E_F=0.014$, we obtain the distribution of
the emerging robust helical surface states in Fig.5(d), and it has been integrated along $z$ axis.
As we can see, the red line and blue line seem degenerated, which originates from
the coupling between the emerging robust helical surface states on $y=0$ and $y=20a$.
Next, a little amount of magnetic impurities in x direction ($M_x\sigma_x\otimes\tau_0$) are doped on $y=0$.
In Fig.5(b), we find that one of the two pairs of emerging robust helical surface states opens
a small gap, while one of the two pairs of gapped normal surface states tends to close.
In this case, the only existing gapless surface states are the emerging robust helical ones
concentrating on $y=20a$ as shown in Fig.5(e).
At last, when the strength of magnetic impurities is strong enough,
such as Fig.5(c) with $M_x=0.9$, there exist
two pairs of gapless surface states consisting of one pair of emerging robust helical surface states and one pair of normal surface states.
The probability density distribution in this case is shown in Fig.5(f).
The blue and red lines represent the emerging robust helical surface states and the normal ones, respectively.
We find only the surface states concentrating around $y=20a$ exist now.
In fact, every characteristic in Fig.5(a)-(f) of Cd$_3$As$_2$ owns its correspondence in Fig.1(a)-(f) of 2D BHZ model,
and the physical pictures between these two models are exactly the same.

\begin{figure}[h]
\includegraphics [width=\columnwidth, viewport=0 0 879 338, clip]{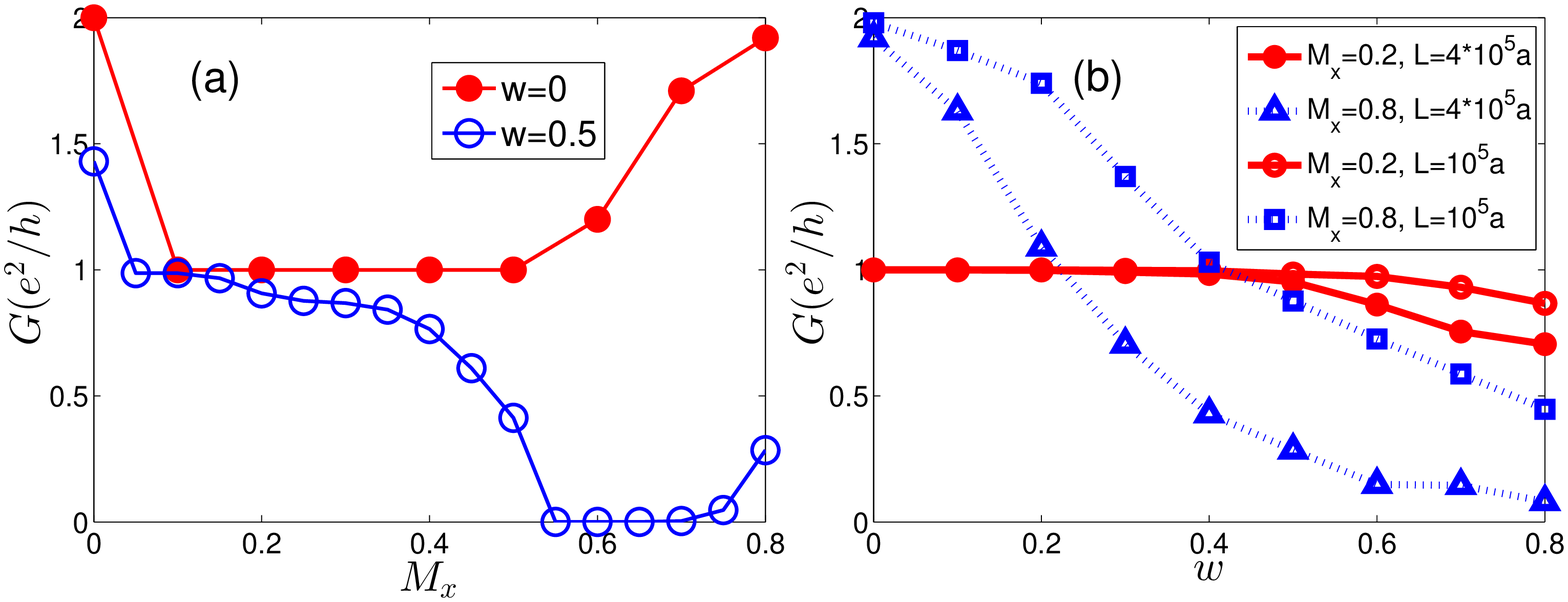}
\caption{(Color online)
(a): The red and blue lines stand for the conductance $G$ vs $M_x$ in Cd$_3$As$_2$ for two Anderson disorder strengths $w=0$ and $w=0.5$, respectively.
(b): The red and blue lines stand for the conductance $G$ vs $w$ in Cd$_3$As$_2$ for two fixed magnetic impurity strengths $M_x=0.2$ and $0.8$,
respectively.
}\label{bulkbandsevolution}
\end{figure}

Finally, we also calculate the conductance of Cd$_3$As$_2$ with Anderson-type disorder in a two-terminal system.
The Fermi energies of the two leads and the central part are $E_L=E_R=0.5$ and $E_C=0.01$, respectively.
Without loss of generality, the disorder is only added on the $y=20a$ boundary.\cite{footnote}
Fig.6(a) tells us how the conductance varies with the strength of magnetic impurities $M_x$.
In the red line, there is no Anderson disorder and the distance between the two leads is $4\times10^5a$.
Similar to 2D BHZ model, the conductance
falls to a plateau of $G=e^2/h$ from the beginning
$G=2e^2/h$ rapidly, and then goes back to $G=2e^2/h$. However, if the disorder
strength is chosen $w=0.5$ shown by the blue line, the conductance decreases from $G=2e^2/h$
to $G=e^2/h$, and never goes back to $G=2e^2/h$ but continues falling to nearly zero.
Similar as the green line in Fig.2(a),
a critical point around $M_x=0.6$ also exists, but exhibits much more explicit than that in BHZ model.
According to the conductance behavior under moderate disorder strength $w=0.5$,
we can claim the two-terminal device is in ``on'' status when $M_x\in [0.1,0.3]$ and in ``off'' status when $M_x>0.55$.
In order to better understand the performance of the topological device,
we also show how the conductance varies with Anderson disorder strength $w$ with fixed magnetic impurity strength $M_x$ in Fig.6(b).
When the device is in the claimed ``off'' status (such as $M_x=0.8$),
the conductance $G$ quickly falls to zero with the increasing disorder strength $w$.
In contrast, when the device is in the claimed ``on'' status (such as $M_x=0.2$),
the conductance nearly maintains at $G=e^2/h$, no matter how the disorder strength $w$ or the length of the strip changes.
According to Fig.6(b), a relatively high on-off ratio about $10:1$ is obtained when $w>0.4$, which is suitable for the topological devices.
Moreover, the stronger the disorder strength is and the longer the stripe is, the higher the on-off ratio can be obtained.
Nevertheless, disorder strength can not be too strong in case of the broken of the ``on'' status.
In fact, Fig.6(b) is consistent with Fig.2(e),
which is also the result of finite-size effect and the key to switch on or off the current
by doping magnetic impurities on one surface.

To summarize, all conclusions obtained from 2D BHZ model can also be found in Cd$_3$As$_2$ type
Dirac semimetals,
which generalizes our proposals building new topological devices to a much broader prospect.

\section{Conclusion}

In conclusion, we show that in $Z_2=0$ topological systems, one can
switch on and off the current nonlocally just by doping magnetic impurities on one edge or surface of the sample,
which enables us to build topological devices through emerging robust helical surface states
in $Z_2=0$ topological systems.
This proposal is first demonstrated in 2D $Z_2=0$ anisotropic BHZ model and 3D Wilson-Dirac type model in detail.
Notably, it is also generalized to realistic Cd$_3$As$_2$ type Dirac semimetals.
Since the manipulation methods proposed here are all nonlocal ones, $Z_2=0$ topological system
with emerging robust helical surface states
exhibits much brighter prospects in building topological devices than the traditional $Z_2=1$ STI.

\section*{ACKNOWLEDGMENTS}
This work was financially supported by MOST of China (Grant No. 2012CB821402), NBRPC (Grant No. 2014CB920901),
NSFC (Grants Nos. 91221302, 11374219, 11204065(JTS), and 11474085(JTS)),
and the NSF of Jiangsu province BK20130283.


\begin{thebibliography}{10}
\bibitem{footnote}
Since the boundary contacts the outside environment directly,
the disorder on the boundary should be stronger than that in the inner space.
Moreover, in 2D BHZ model, we have demonstrated that the proposed effect exists even though the disorder is added in the whole system.
In Cd$_3$As$_2$ Dirac semimetal materials, we also found the existence of this effect,
and we just consider this approximate condition to make the effect more explicite.
\end{thebibliography}
\end{document}